# Terahertz control in a transmission electron microscope

J. Kuttruff[†], D. Nabben[†], A. C. Zimmermann, A. Ryabov*, P. Baum*

*Universität Konstanz, Fachbereich Physik, 78464 Konstanz, Germany*
[†]*These authors contributed equally to this work*
*andrey.ryabov@uni-konstanz.de
*peter.baum@uni-konstanz.de

**Abstract:** Ultrafast electron microscopy provides a movie-like access to structural dynamics of materials in space and time, but fundamental atomic motions or electron dynamics are, so far, too quick to be resolved. Here we report the all-optical control, compression and characterization of electron pulses in a transmission electron microscope by the single optical cycles of laser-generated terahertz light. This concept provides isolated electron pulses and merges the spatial resolution of a transmission electron microscope with the temporal resolution that is offered by a single cycle of laser light. Central to these achievements is a perforated parallel-plate metallic waveguide in which transverse velocity mismatch and magnetic forces are mitigated by electrically constructive and magnetically destructive interferences of incoming and reflected terahertz half-cycles from a displaced waveguide termination. Measurements of spatial chirp via energy-filtered imaging reveal flat pulses with no transversal deflection or temporal aberrations at the specimen. We also report the all-optical control of multi-electron states and discover a substantial two-electron and three-electron anti-correlation in the time domain. These results open up the possibility to visualize atomic and electronic motions together with their quantum correlations on fundamental dimensions in space and time.

**Introduction**

Ultrafast transmission electron microscopy [1-4] combines the time resolution of pulsed laser systems or optical cycle periods with the spatial resolution of an electron microscope for investigating non-equilibrium dynamics in complex materials and nanostructures. For example, ultrafast electron microscopy can track the motion of atoms in structural transitions in solids [5-8], probe magnetic pinning effects [9,10], visualize plasmon-polariton and phonon-polariton dynamics [11-13], understand optical vortex annihilation [14] or image electromagnetic fields [15-19] and sub-cycle waveforms [20,21]. However, the time resolution in such pump-probe experiments is limited by the duration of the laser-generated electron pulses. Rapid electron pulse sequences and pulse trains are useful for measuring cycle-reversible dynamics [4], but compression of isolated electron pulses remains to be achieved. Currently, the electron pulses in state-of-the-art electron microscopes are more than 200 fs long [2,3], limited by the unavoidable bandwidth of ultrafast photoemission [22] in conjunction with quantum-mechanical wavepacket dispersion in free space [23]. While a time resolution of 200 fs might be sufficient for some slower motions of heavy atoms [5-8], the observation of fundamental atomic and electronic processes [24] requires a ten to hundred-fold improvement.

**Results**

Figure 1a shows the concept of our experiment. The idea is to use a single optical cycle of terahertz light to control the electron pulses in space and time [25] in a way that preserves the spatial resolution of the microscope. A femtosecond laser (Carbide, Light Conversion) delivers 200-fs long laser pulses at a centre wavelength of 1030 nm. These laser pulses are frequency-doubled and focused on the field-emitter electron source of our transmission electron microscope (JEM F200, JEOL), yielding electron pulses with a duration of around 200 fs at an energy of 80 keV. Terahertz single-cycle electromagnetic fields are generated via optical rectification in lithium niobate [26,27]. The centre wavelength of $\lambda_{THz} \approx$ 1 mm and centre frequency of $\nu_{THz} \approx$ 0.3 THz provide a half-cycle period of 1.7 ps which is large enough to cover the incoming electron pulses in the time domain.

The inset of Fig. 1a shows a phase-space representation of the intended electron control. Initially, the photoemitted electron pulses (red) have an energy width of 0.6 eV, determined by residual mismatch between emitter work function and photon energy [23], and a pulse duration of ~200 fs, determined by the finite emission time. The dispersion of vacuum for non-relativistic electrons further lengthens these pulses during subsequent propagation to ~300 fs (yellow). Next, we use the electric field gradient of a terahertz pulse (violet) for accelerating the trailing part and decelerating the leading part of the incoming electron phase space (yellow). The sign of the dispersion is inverted and the energy spectrum is broadened at the same time (violet arrows). By further free-space propagation, the negatively chirped electron pulses (green) then temporally compress (blue) at a chosen position $f_{temp}$ that is adjusted, for example, to coincide with the specimen location. There, the electron pulses can become as much shorter in time as they broaden in the energy domain [25].

However, the electron beam in our transmission electron microscope has a hundred to a thousand times better beam quality (emittance) than the millimetre/micrometre-sized electron beams in all previous terahertz control experiments [25,28-37]. We are therefore extremely sensitive to potential aberrations and unwanted phase-space couplings that often originate, for example, from magnetic-field phenomena [38] or inhomogeneous acceleration effects [25]. In addition, the ability of a transmission electron microscope to achieve atomic resolution requires large magnetic lenses, cooling water, x-ray shielding and magnetic isolation from the environment. The associated construction principles limit the available numerical aperture for free-space terahertz focusing to about $10^{-2}$. The achievable terahertz power density at such a focus is at least two orders of magnitude too small for electron pulse compression, calling for an alternative approach.

Figure 1b depicts the principle of our terahertz-electron interaction geometry. We apply a metallic parallel-plate waveguide (see Supplementary Fig. 1) that is in principle lossless and dispersion-free [39]. The total length is 180 mm. A taper at the end of the waveguide confines the electric field down to 100 μm or one tenth of the terahertz wavelength in longitudinal direction ($z$-axis), providing at the same time the necessary sub-cycle field-electron interaction [25] and an electric field enhancement of ~20 with respect to a free-space focus. A sub-wavelength hole that is drilled through the waveguide walls allows electrons to propagate through and interact with a nearly time-frozen terahertz field. In order to achieve electron pulse compression without spatial aberrations and emittance growth, we need spatially homogeneous and time-varying electric fields in longitudinal direction but no magnetic fields that would deflect the electron beam into a transverse path. On the other hand, for streaking by sideways deflection [25], we ideally need time-dependent magnetic fields but no electric components. We solve these requirements by shorting the waveguide with a metallic end wall at variable distance to the interaction region. The resulting formation of a terahertz standing wave is depicted in Fig. 1b. Due to the $\pi$ phase jump of the electric field component upon reflection from the metallic end (violet arrow), the field nodes of $E(x,t)$ (red) and $B(x,t)$ (blue) are offset by one quarter of the terahertz carrier wavelength, allowing to enhance either the electric fields (for acceleration and deceleration) or the magnetic fields (for transversal deflection), depending on the hole position at locations ① or ②. For electron pulse compression (①), we place the electron beam (dotted line) at $\lambda_{THz}/4$ away from the back wall, where the electric fields (red) are amplified but the magnetic fields (blue) are cancelled out. For delay-dependent deflection for pulse characterization (②), we can place the electron beam (dotted line) to a magnetic-field antinode (red) at $\lambda_{THz}/2$, where there remain only the amplified magnetic fields (blue).

Figures 1c-d show snapshots of the simulated electric and magnetic field distributions inside the waveguide for our experimental geometry (see Materials and Methods). For the case of pulse compression (①), we see that the longitudinal *E*-field is maximally enhanced at the position of the electron beam (blue arrow), providing longitudinal energy modulation as a function of terahertz delay. At the same position, we see that the magnetic fields indeed mostly cancel out. There remains only a cylindrically symmetric swirling *B*-field around this point, necessary by the Panofsky-Wenzel theorem [40]. Like in microwave cavities [41], we expect that the associated spatial focusing will be weak and easily compensated by the microscope's available electron optics. For the case of sideways streaking (②) with the termination at $\lambda_{THz}/2$, we see an *E*-field minimum and a *B*-field maximum that should provide a pure time-dependent sideways deflection (blue arrows) at no substantial longitudinal effects. The function of our multi-purpose structure is hence selected by choosing appropriate sub-cycle terahertz delays in conjunction with different electron beam positions relative to the end wall of the waveguide for temporal pulse compression, transversal streaking and deflection, transient lensing, controlled electron acceleration/deceleration or combinations thereof.

Figure 2a shows a measurement of the time-dependent energy gain or loss that the electron pulses obtain in a compression-optimized waveguide structure. We plot here the measured electron energy spectrum as a function of the terahertz-electron delay. We see that the energy change follows the terahertz waveform, that is, the acceleration or deceleration of the electrons is directly proportional to the approximately time-frozen electric field at the electric anti-node (compare Fig. 1b). The number of field cycles is about two times larger than the incoming single-cycle electric waveform, because the reflected and incoming main field peaks interfere with each other at $\frac{1}{2}\lambda_{THz}/c$, where c is the speed of light in the waveguide. We also see some residual waveguide dispersion, that is, a shorter terahertz wavelength at early times and a longer wavelength at later times. This effect arises from confinement of the waveguide to ~4 mm in the *y*-axis for mechanical reasons, violating lateral invariance [39]. A Fourier transform of the time-domain data is shown in Supplementary Fig. 2.

Figure 2b shows an analysis of the time-dependent energy gain/loss (black curve) in comparison to the measured transversal deflection of the electron pulses (green) by the magnetic parts of the terahertz wave. Between -4 ps to around -1 ps, the sideways electron deflection by magnetic effects follows directly the oscillations of the energy gain/loss by the electric fields. Here, there is no reflection yet, and we simply see the plane-wave-like incoming terahertz pulse with $E(t)$ proportional to $B(t)$. Between -1 ps and 1 ps, gain/loss and deflection get out of phase by approximately 90°, because a reflected pre-cycle now interferes with the incoming main cycle, producing a mixture between a standing and a travelling wave. Finally, for terahertz time delays above 1 ps, the transverse beam deflections vanish (flat green curve) and there remain solely a pure energy gain or loss (oscillating black curve). We see that the delay time A is ideal for pulse compression while, for example, a delay B would introduce compression and deflection at the same time, because the necessary standing wave has not had time to emerge. Delay C would be ideal for intentionally stretching the electron pulses in time.

In addition to avoiding magnetic deflection effects, it is also essential to provide a homogeneous acceleration or deflection over the entire electron beam profile. Our waveguide provides this uniformity by the absence of travelling-wave effects in combination with electron transmission holes that are almost ten times smaller than $\lambda_{THz}$. In order to confirm this prediction, we invoke energy-filtered transmission electron microscopy (EFTEM) and measure the position-dependent energy distribution of the compressed electron pulses (delay time A) at the specimen location ($f_{temp} \approx 180$ mm). Figure 2c shows in the left panel the position-dependent mean energy change. We see a flat profile that is almost independent of location within the beam. Consequently, there are neither spatio-energetic correlations nor tilted electron pulses at the specimen. In contrast, Fig. 2d shows in the left panel the same measurement but now for an electron-terahertz interaction at position B where there are remaining travelling waves that pass gradually over the electron beam. We now measure a strong dependence of the electron energy distribution as a function of position in the beam, corresponding to tilted electron pulses at the specimen [38]. Numerical simulations confirm all measurements; see the right panels in Fig. 2c-d. In summary, these results show that the reported waveguide concept can indeed provide a uniform and almost aberration-free interaction between beam electrons and terahertz light in sub-cycle times.

**Electron Pulse Compression**

We now report electron pulse compression and characterization directly in the time domain. To this end, we employ two of our shorted metallic terahertz waveguides in sequence, one for electron pulse compression by the electric anti-node and one at the specimen location for temporal pulse characterization by the magnetic fields.

We first measure the temporal pulse profile of the uncompressed electron pulses as generated from laser photoemission. Figure 3a shows a series of deflected electron beams, called a deflectogram, as a function of the terahertz delay at the streaking waveguide. The mean deflection follows the magnetic part of the terahertz waveform. At the strongest magnetic gradient (dotted line), the deflection is a steep function of the terahertz time delay, and the streaked beam width relates to the incoming electron pulse duration because electrons at different arrival times are deflected into different directions at a given delay time. We measure a streaking speed of $\frac{dx}{dt}$ =0.11 px/fs, corresponding to a temporal resolution of the streak camera of ~13 fs (see Materials and Methods). Figure 3b shows a zoom-in of the deflectogram at the delay of maximum streaking speed (dotted line). In comparison, Fig. 3c shows the streaking data in case that the temporal compression in the first waveguide is switched on at optimum compression strength. We see a substantial sharpening of the curve, related to a substantial decrease of electron pulse duration.

Cuts of the measured deflectograms at $t$ = 0 fs are shown in Fig. 3d and Fig. 3e for the uncompressed and compressed electron pulses, respectively. In order to determine the electron pulse duration in both cases, we consider the measured streaking speed and the measured widths $\Delta x$ of the deflectograms at zero delay. When neglecting the finite width of the unstreaked electron beam due to imperfections of the source, the electron pulse duration is given by $\Delta t = \Delta x / \frac{dx}{dt}$ [37,38]. We obtain electron pulse durations of 270 fs for the electron pulses directly from the source (see Fig. 3d) and 19 fs for the compressed electron pulses (full widths at half maximum). In practice, the measured profiles of Fig. 3d-e are convolutions of the genuine streaking dynamics with the unstreaked electron beam width on the detector, limited by beam emittance (see Materials and Methods). This effect is here not taken into account, so our reported pulse durations are upper bounds. In contrast to shorter pulses that have been achieved before in the pulse-train regime [4], these pulses are isolated and therefore applicable to pump-probe investigations of non-equilibrium materials.

**Spatial Resolution**

To showcase the ability to obtain transmission electron microscopy images and diffraction patterns with our all-optically compressed electron pulses, we report in Fig. 4 a series of comparisons of microscopy data obtained with or without terahertz pulse control. Figure 4a shows in the left panel a microscope image of gold nanoparticles in comparison to the same object imaged with our 19-fs electron pulses. Figure 4b shows line cuts of the data at the indicated positions (dotted lines). We see that spatial resolution is not impaired and feature sizes below 3 nm can be resolved, probably limited by the roundness of our specimen. Figure 4c shows in the left panel a diffraction pattern obtained from a (100) silicon crystal and in the right panel a measurement with our 19-fs electron pulses. We see Bragg spots up to 6th order at no discernible differences of intensity or shape. Figure 4d shows the measured electron spot size at the specimen as a function of the objective lens defocus. The resulting waist scans for the uncompressed electron pulses (orange) and for the 19-fs pulses (green) are almost identical with only a constant displacement of about 20 μm. This defocus corresponds to an effective negative lens in the terahertz compressor of about -360 mm which agrees with expectations from microwave cavities in $TM_{010}$ mode where the spatial focus distance is $f_{spatial} = -2f_{temp}$ [41]. Notably, this weak defocusing action of our waveguide originates from transient circular magnetic fields that do not obey the Scherzer theorem and are aberration-free [41]. Terahertz-compressed electron pulses have therefore no more spherical aberrations than a conventional electron beam. It is also possible to use our terahertz modulators to monochromatize a dispersed pulsed electron beam [42], to chop short narrow-band electron pulses out of longer ones [43] or to create almost any other desirable phase-space distributions for novel imaging techniques.

**Time-Energy Phase Space**

According to Liouville's theorem (see inset of Fig. 1a), any electron pulse compression must be associated with a corresponding energy bandwidth gain. Figure 5 shows a measurement of the energy spectrum of our uncompressed electron pulses (black) in comparison to the energy spectra of our terahertz-controlled electron pulses at three distinct terahertz delays (green, blue, orange). For the conditions of pulse compression at a rising slope of the electric anti-node (see Fig. 1b), we see a broadening from ~0.6 eV (black) to ~55 eV (green). If we assume conservation of time-energy phase space (see inset of Fig. 1a) and no aberrations, the incoming 270-fs long electron pulses could in principle be compressed to 3-fs electron pulses at the specimen, as a lower limit. From the reported direct time-domain characterization (Fig. 3e) we can therefore conclude that our compressed electron pulses are longer than 3 fs but shorter than 19 fs in our experiment. For applications in which chromatic aberrations are critical, for example scanning transmission electron microscopy or sub-atomic spectroscopy [44], we recommend to use aberration corrector systems [45] or minimize the initial time-energy phase space volume by using shorter laser photoemission pulses and a photon energy close to the emitter work function [23].

The orange and blue traces in Fig. 5 show the measured energy spectrum of the electron pulses at negative and positive turning points of the electric fields at the electric anti-node. We see that the incoming electron can be accelerated or decelerated by ±100 eV. Although the incoming electron pulses now arrive at a crest and not at a slope of the terahertz wave, the waveguide still provides destructive magnetic interference that cancels out any magnetic phenomena (see Fig. 1b). The accelerated or decelerated electrons therefore remain on axis and can be used, for example, for an all-optical adjustment of electron arrival times, for characterization or manipulation of correlated electron pairs [46,47] or for coherent electron energy loss interferometry experiments [48].

**Multi-Electron Phase-Space Control**

As a first application of electrons under single-cycle light control, we demonstrate several manipulations of the multi-dimensional phase space of Coulomb-correlated multi-electron states. In particular, we produce two-electron and three-electron packets [46,47] and use terahertz cycles to measure and amplify their energetic and temporal correlations. To this end, we use a single-electron counting camera [49] to produce streaking data as a function of the measured number of electrons per pulse. While multi-electron states are known to have an eV-scale anti-correlation in the energy domain [46,47], their temporal properties are currently unclear.

Figure 6a shows the measured terahertz streaking spectrogram for electron states with exactly one electron per pulse ($N=1$). In contrast, Fig. 6b shows the same data for exactly two-electron states ($N=2$). We see the emergence of two separated curves at substantial time delay. The energetic separation is strongest at the terahertz slopes (-0.9, 1.1 ps) and minimal at the turning points. Figure 6c shows the measured single-electron and two-electron arrival times, obtained by integration along the black rectangle of Fig. 6a. Indeed, we see one peak for $N=1$ (black) and two separated peaks for $N=2$ (blue) in which the anti-correlation gap matches the arrival time of the single-electron states. The width of the peaks relates to the direct electron pulse duration for the $N=1$ case and the effective individual single-electron pulse durations in the $N=2$ case. The temporal distance of electrons 1 and 2 in the pair state is ~290 fs. Figure 6d shows a plot of the pair density as a function of the measured arrival times $T_1$ and $T_2$ for a fixed terahertz time delay of ~1.1ps. We see a clear anti-correlation within the pair state with a strong gap at zero timing difference $T_1 - T_2$. Whenever electron 1 is early, electron 2 is late, and vice versa. The two electrons almost never arrive at the same time, that is, within <50 fs (white gap in Fig. 6d). The two extra spots at ±380 fs are artefacts from nonlinearities of the applied terahertz single-cycle at extreme arrival times.

In order to understand these results, we argue that more than one electron cannot be emitted along the optical axis from a nanometre source at exactly the same time. If two electrons are emitted with a small delay, the mutual Coulomb forces quickly induce an anti-correlation of forward momentum [46,47] that subsequently converts by the dispersion of free space for our electrons into the measured anti-correlation in the time domain. Three-electron states also show similar effects, as we show in Supplementary Fig. 3.

We can also use our terahertz control to amplify the initial few-eV energetic anti-correlation in multi-electron states [46,47] about tenfold in the energy domain. We let the pulses disperse for ~6 ns and then use terahertz acceleration/deceleration by an electric anti-node to tilt the phase space from the temporal into the energy domain. Figure 6e shows a histogram of the measured electron energy differences $E_1 - E_2$ in the resulting pair density state at maximum acceleration gradient. We see two peaks at symmetric locations around ±15 eV and a gap at zero energy. We thus increase their energy separation by one order of magnitude as compared to the initial separation at the source. These observations show that we can not only control and shape single-electron pulses in space and time but also use our all-optical control to modify the multi-dimensional properties of multiple particles by the optical fields of single-cycle light. If multi-electron states will eventually be prepared at enough coherence to become quantum-mechanically entangled, for example in the time-energy or position-momentum domains, the here reported concepts will be available for characterization and quantum state control.

**Discussion and Outlook**

The combined results show that terahertz-electron interaction in node-matched waveguides is a versatile and capable tool for manipulating the phase space of electron pulses in a transmission electron microscope. The reported ability to compress electron pulses from hundreds to tens of femtoseconds advances ultrafast electron microscopy from the domain of rate-driven dynamics [6,8] and effective media [5,6,8] into the domain of atomic motions and phonon dynamics on fundamental dimensions in space and time. With plausible improvements of initial emission times and bandwidths at the source, the reported phase space reshaping concepts have potential to create electron pulses approaching the attosecond frontier [50] and advance attosecond electron microscopy [4] from the pulse-train into the isolated-pulse regime. More generally, a transmission electron microscope with its electron beam under control of the optical electric and magnetic fields of engineered single-cycle light will be a versatile and groundbreaking instrument for material science, nanophotonics, metamaterials, electron quantum optics and related research.

**Material and Methods**

**Waveguide structures:** Terahertz waveguides are CNC-machined (computerized numerical control) from copper or brass and afterwards polished to close-to-optical quality (see Supplementary Fig. 1). The total length is 180 mm and the entrance aperture is 4 mm in transversal direction ($y$ axis) times 4 mm in longitudinal direction ($z$ axis). A first tapering reduces the profile to the dimensions of the parallel-plate waveguide ($4 \times 2$ mm$^2$). Within the last 15 mm, the waveguide shrinks to 600 μm (transversal)×100 μm (longitudinal). A hole with diameter of 150 μm allows for electron transmission through the waveguide. The entrance of the structure is vacuum-sealed with a 2-mm thick quartz window.

**Experimental details:** Experiments are performed using a transmission electron microscope (JEM F200, JEOL) and a 40-W femtosecond laser (Carbide, LightConversion) at 200 kHz repetition rate. Part of the fundamental laser pulses are used for second harmonic generation (SHG) in a beta barium borate crystal (BBO). The resulting 40-nJ, 515-nm pulses are focused on the emitter with a 300-mm lens, yielding a spot size of ~20 μm in the focus. The remaining laser power is divided and guided to two independent terahertz generation systems that independently drive electron pulse compression and streaking, respectively. In each arm, terahertz single-cycle pulses with centre frequency of 0.3 THz are generated in lithium niobate using a Cerenkov geometry with silicon output couplers [26,27] and focused into the waveguide structures using 100-mm lenses made of polymethylpentene (TPX). The measured terahertz beam diameter in the focus is ~2 mm and therefore smaller than the entrance aperture of the waveguides. From the measured electron energy modulation (Fig. 2a), we can estimate the terahertz electric field amplitude at the end of the waveguide to be ~$0.9 \times 10^6$ V/m. Energy-filtered electron microscopy images (Fig. 2, Fig. 6) are generated with a post-column electron energy spectrometer (CEFID, CEOS). Electrons are detected at negligible noise level using an event-based direct electron detector (Timepix3, Amsterdam Scientific Instruments) and converted to single-electron events with a fast flow-counting algorithm [49].

**Streaking resolution:** The time resolution of our streak camera (Fig. 3) is determined by the convolution of the maximum available streaking speed (0.11 px/fs) with the unstreaked electron spot size on the camera [38]. Using our pixelated direct electron detector, we can focus electrons down to spot sizes of ~1.5 px, limited by the detection physics of the device [49]. Consequently, the estimated temporal resolution of the streak camera is ~13 fs. Due to a slight misalignment of the streaking waveguide, there are residual magnetic effects that cause a time-dependent focusing or defocusing in addition to the intended time-dependent deflection. The streaking data therefore develops a delay-dependent width by cylindrical-lens effects. Consequently, all measured streaking widths and pulse durations are upper bounds [38]. Although we optimize the electron beam focus around the steepest slope (dotted line in Fig. 3c), in order to minimize the spatial defocus and maximize the time resolution, the actual electron pulses could be shorter than we state (19 fs).

**Simulations:** Time-dependent electromagnetic fields and electron phase spaces are calculated with a finite-element solver that is coupled to a particle-in-cell simulator (CST studio suite, Dassault Systèmes). For simulating the fields, a Gaussian-profiled plane wave with radius of 1 mm at center frequency of 0.3 THz is coupled into the waveguide. Electron pulses of 270 fs duration and 80 keV energy then interact with the full, time-dependent electromagnetic field in the waveguide and we extract position, velocity and time of each electron. Further propagation in free space is simulated using relativistic equations of motion. For results in Fig 2 c-d, the obtained energy-position data is scaled down to match the experimental beam size. Even shorter isolated electron pulses can be obtained from sharper electron emitters [2] and optimized photoemission wavelengths [23]. Indeed, simulations with 100-fs input pulses (see Supplementary Fig. 4) show the emergence of isolated 8-fs electron pulses. Much shorter pulse durations can be obtained at shorter focal distances or if higher electron acceleration voltages are applied [50].

**Data availability:** The data supporting the findings of this study are available from the corresponding author upon reasonable request.

**Acknowledgements:** This research was supported by the Dr. K. H. Eberle Foundation and the German Research Foundation (DFG) via SFB 1432.

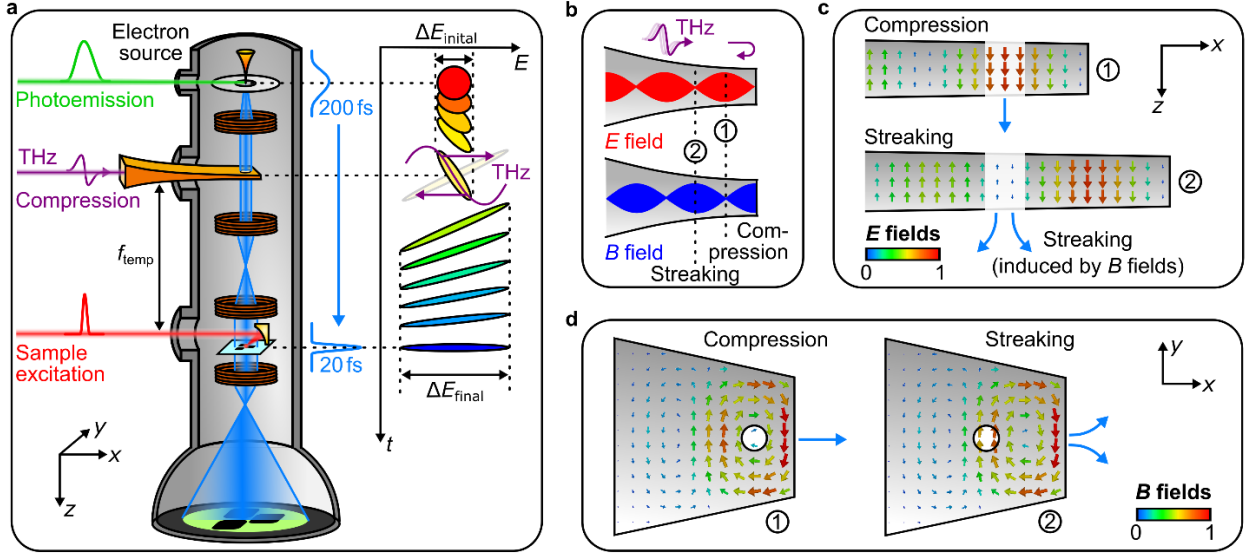

**Fig. 1.** All-optical electron beam control in a transmission electron microscope. **a,** Experimental setup and pulse compression concept. A photoemission laser (green) creates short electron pulses (blue) that intersect with the electric or magnetic anti-nodes of a metal waveguide (orange) under terahertz illumination (violet). A pump laser or secondary terahertz beam (red) excites a material and pump-probe or streaking images are measured on a screen (green). Inset: Phase space diagram of terahertz compression in the time-energy domain. If there are no space charge effects [23] and all forces are sufficiently linear in time [25], the initial electron pules (red) become as much shorter in time as they broaden in the energy domain (blue). **b,** Generation of electric and magnetic anti-nodes by reflection of terahertz pulses at a displaced waveguide end. The purely electric (red) and purely magnetic (blue) anti-nodes are ideal for pulse compression or streaking of electron beams at positions ① or ②, respectively. **c,** Simulated longitudinal electric fields inside the waveguide at ideal time delays for compression (upper part) and streaking (lower part). **d,** Simulated in-plane magnetic fields at ideal time delays for compression (left part) and streaking (right part). The circle depicts the sub-wavelength entrance and exit holes for the electron beam.

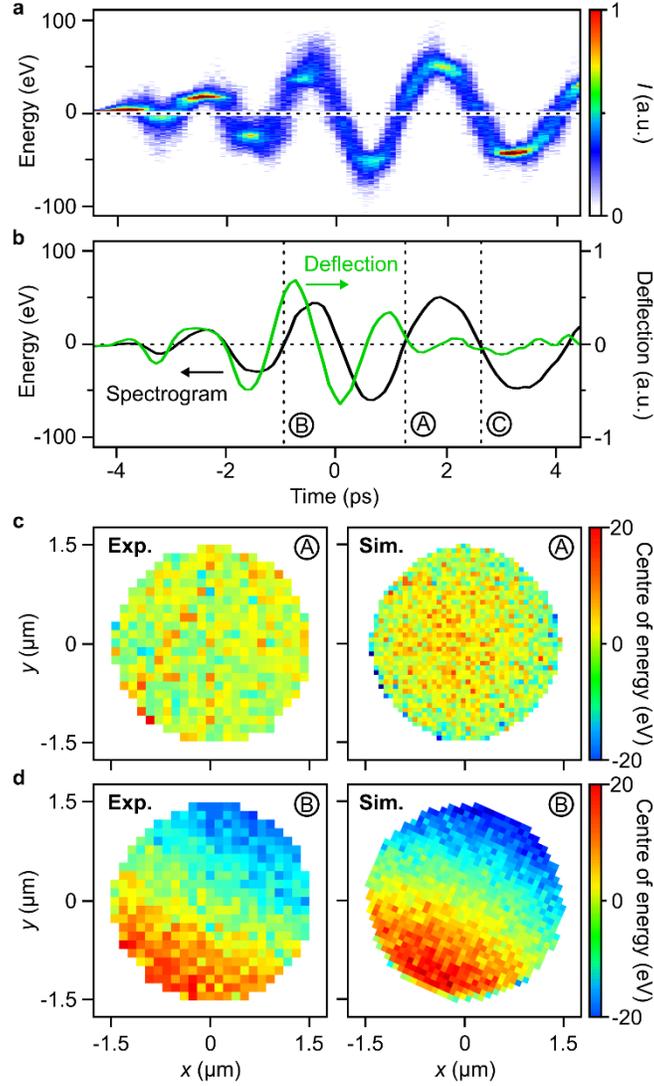

**Fig. 2.** Energy modulation and standing-wave terahertz interaction. **a,** Energy spectrum of electron pulses after interaction with the compression waveguide as a function of terahertz delay. **b,** Measured energy modulation (black) by electric fields in comparison to the measured transversal deflection by magnetic fields (green). Time A is ideal for pulse compression, time B produces tilted pulses and time C is ideal for stretching electron pulses in time. **c,** Measured (left) and simulated (right) central energy modulation of compressed electrons (A) as a function of position $x$ and $y$ across the beam. **d,** Measured (left) and simulated (right) central energy modulation of the electron beam at an improper interaction time (B).

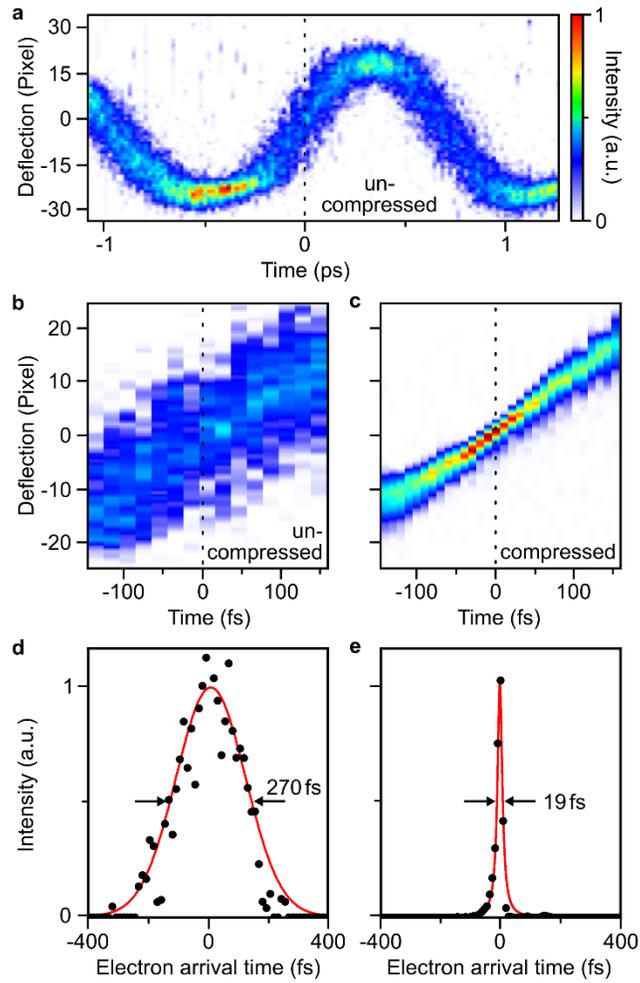

**Fig. 3.** Characterization of electron pulses by streaking and generation of 19-fs pulses. **a,** Terahertz streaking curve of uncompressed electron pulses. **b,** Streaking data at the steepest slope. **c,** Streaking data for electron pulses after terahertz compression. **d,** Measured time-dependent electron current for uncompressed electron pulses and Gaussian fit (red). **e,** Measured time-dependent electron current for compressed electron pulses and Lorentzian fit (red).

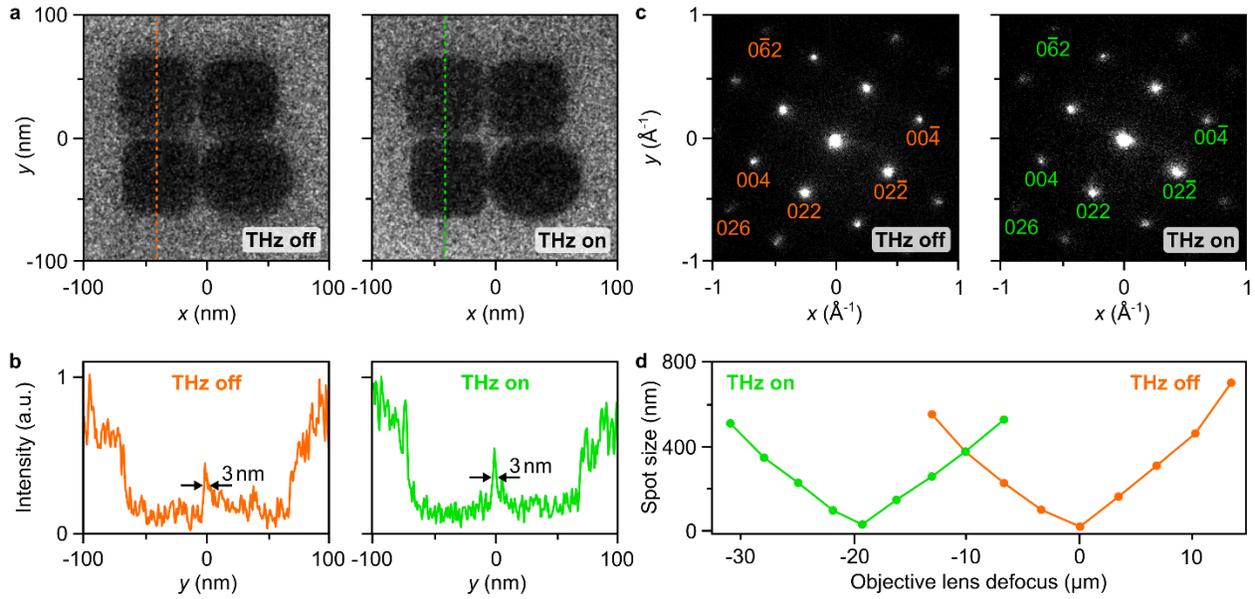

**Fig. 4.** Transmission electron microscopy and diffraction with terahertz-compressed electron pulses. **a,** Microscope images at a magnification of 200,000 of gold nanoparticles without (left panel) and with (right panel) terahertz compression. **b,** Line cuts at the dotted lines for uncompressed (orange) and compressed 19-fs electron pulses (green). Features sizes of 3 nm can be resolved in both cases. **c,** Diffraction pattern of crystalline silicon recorded without (left panel) and with (right panel) terahertz compression. **d,** Measured beam sizes of the uncompressed (orange) and compressed (green) electron pulses as a function of the microscope objective lens defocus. The focus shifts by 20 μm.

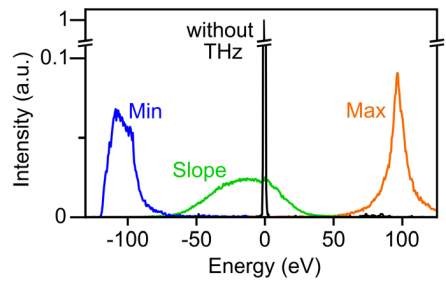

**Fig. 5.** Electron acceleration and deceleration. Measured electron energy spectra for initial electron pulses (black), maximum acceleration (orange), maximum deceleration (blue) and pulse compression (green).

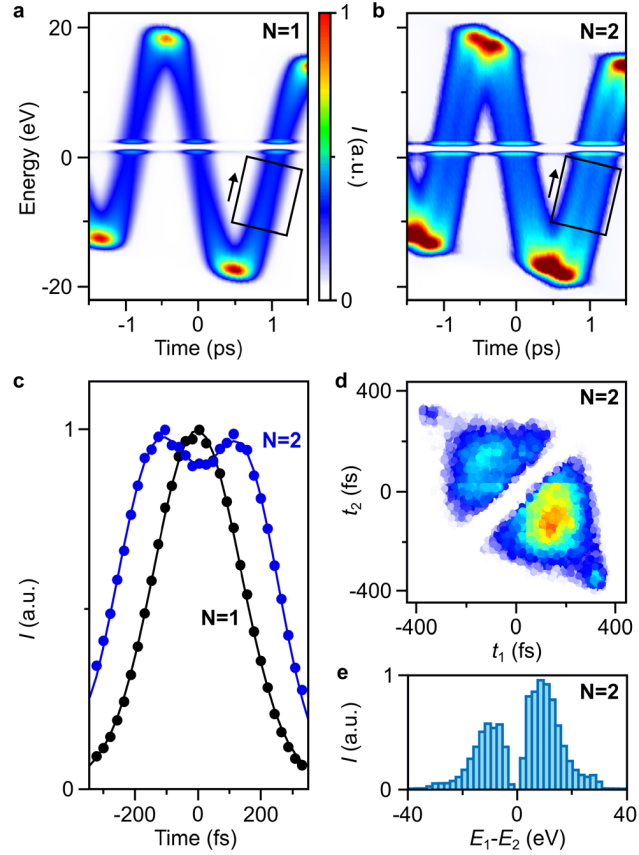

**Fig. 6.** Multi-electron phase-space characterization and control. **a,** Measured spectrogram for one-electron states ($N = 1$). **b,** Measured spectrogram for two-electron states ($N = 2$). Black rectangles, regions of interest for our further analysis; white lines, artefacts from our detector system. **c,** Measured electron arrival times for exactly one electron (black) and exactly two electrons per pulse (blue). **d,** Measured pair density as a function of arrival times $t_1$ and $t_2$ for electrons 1 and 2 within a pair . Data is recorded at terahertz time delay of 1.1 ps. **e,** Measured histogram of energy differences $E_1$-$E_2$ within a pair state after terahertz-driven energy magnification. The initial separation of ~2 eV [46,47] becomes ~15 eV.

**Supplementary Materials**

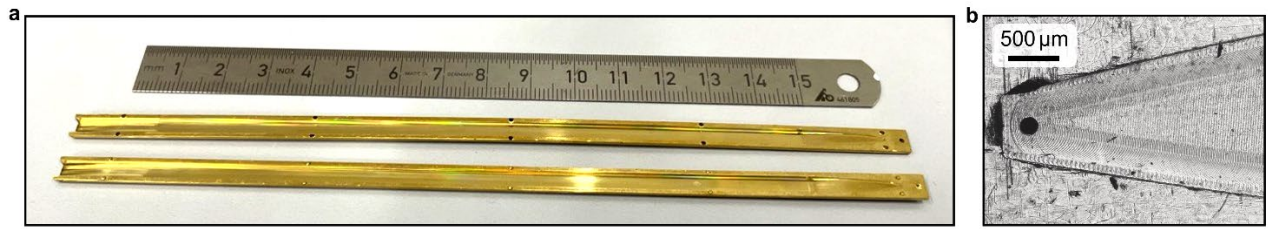

**Supplementary Fig. 1.** Fabricated waveguide structures. **a,** Photograph of the upper and lower half. **b,** Optical microscope image of the terahertz-electron interaction region.

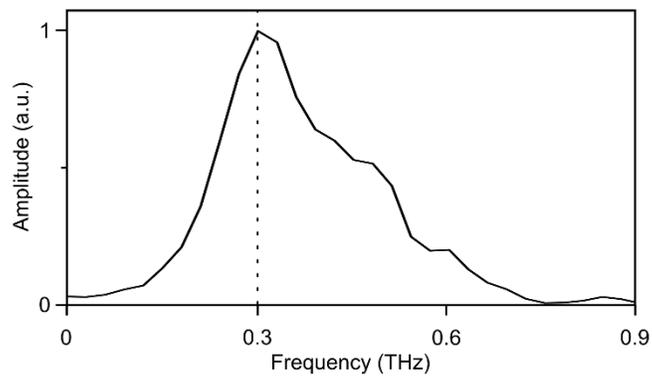

**Supplementary Fig. 2.** Terahertz spectrum obtained by Fourier transformation of the measured electron energy modulation as shown in Fig. 2a.

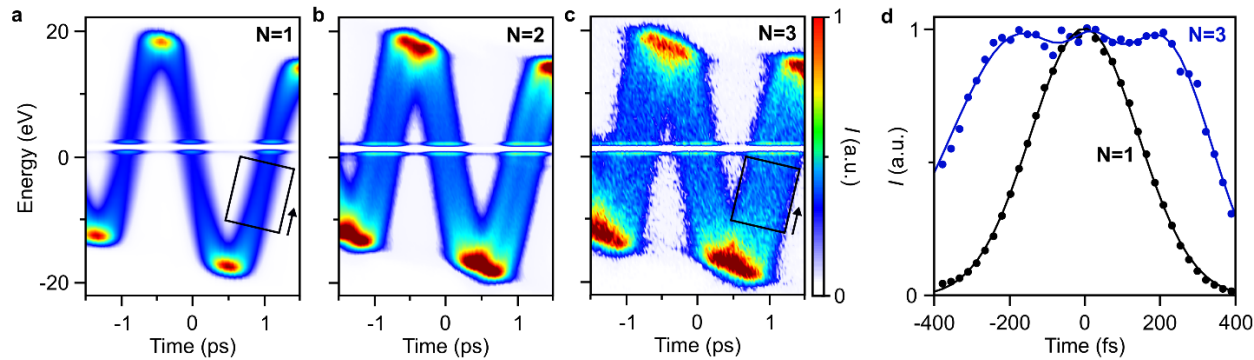

**Supplementary Fig. 3.** Three-electron correlations. **a-c,** Measured energy spectrograms for one-electron states ($N = 1$), for two-electron states ($N = 2$) and for three-electron states ($N = 3$). Black rectangle, region of interest for further analysis. **d,** Evaluated electron arrival times (sum of the spectrogram along the black rectangle) for $N = 1$ (black) and $N = 3$ (blue). Dots, experimental data; solid lines, Gaussian fits.

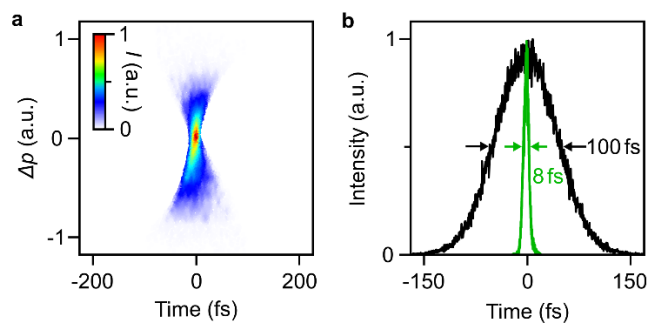

**Supplementary Fig. 4.** Numerical simulation of terahertz pulse compression for shorter initial pulses. **a,** Resulting longitudinal phase space in the momentum-time domain at the temporal focus for input pulses of 100 fs duration. **b,** Simulated temporal pulse profile before (black) and after (green) terahertz compression.